\documentstyle[twocolumn,aps,prl,epsfig]{revtex}

\begin{document}

\wideabs{

\title{The spectral properties of noncondensate particles in Bose-condensed atomic
hydrogen}

\author{Yu. Kagan and A. E. Muryshev}

\address{Russian Research Center Kurchatov Institute, Kurchatov Square, 123182, Moscow,
Russia.}

\maketitle
\begin{abstract}
The strong spin-dipole relaxation, accompanying BEC in a gas of atomic hydrogen,
determines the formation of a quasistationary state with a flux of particles
in the energy space to the condensate. This state is characterized by a significant
enhancement of the low-energy distribution of non-condensate particles, resulting
in the growth of their spatial density in the trap. This growth leads to the
anomalous reconstruction of the optical spectral properties of non-condensate
particles.
\end{abstract}
}



The discovery of Bose-Einstein condensation (BEC) in a trapped dilute gas of
atomic hydrogen \cite{1,2} has opened a new page in the studies of metastable
gaseous systems at ultracold temperatures. For a long period, trapped atomic
hydrogen remained a very promising system to achieve BEC (see, e.g.,\cite{3,4}).
However, only after the involvement of the forced evaporative cooling mechanism
typical for the works resulted in the BEC discovery in alkali metal vapors \cite{5,6,7}
the achievement of BEC in atomic hydrogen became a reality. 

The experiments reported in \cite{1,2} have brought out a set of specific features
inherent in the hydrogen system. The anomalously low rate of a three-particle
recombination allows one to reach the record density of a Bose condensate of
\( n_{c}\sim 10^{16} \)cm\( ^{-3} \) with the total number of atoms \( N \)
by several orders of the magnitude greater than that in the experiments on alkali
gases. The lightest mass of hydrogen atoms in combination with so high density
determines the superfluid phase transition temperature \( T_{c} \) exceeding
that for the experiments with alkali gases by two orders of the magnitude. 

One of most interesting features of the hydrogen experiments is impossibility
to achieve a relatively high concentration of the condensate fraction. This
result is a direct evidence for the phenomenon of the ``burning of a condensate''
\cite{8} related to a strong increase of the gas density at the condensation
and, as a result, drastic enhancement of the spin relaxation rate (see also
\cite{9}). The low rate of the evaporation cooling due to an extremely small
scattering length \( a\approx 0.65{\rm \stackrel{\circ }{A}} \) makes this
phenomenon responsible for the kinetics of the formation of a condensate in
a gas of atomic hydrogen. This led to the prediction \cite{8} that the concentration
of the condensed fraction cannot exceed several percents. It is worthwhile to
note that the spatial density of the condensate may be large.

In the experiments \cite{1} an unexpected phenomenon has been observed. The
shift of the line of the 1s-2s transition for non-condensate particles is found
to increase abruptly after the BEC transition. This is especially surprising
since the condensate volume is only about \( 1\% \) of the volume of the thermal
cloud and the estimations presented in \cite{1} could not explain the observed
picture. The aim of the present paper is to reveal the nature of this interesting
phenomenon. 

Under conditions of the experiments \cite{1,2} the interparticle collision
time \( \tau  \) is small compared with the lifetime of the system \( \tau _{L} \).
As a result, a quasistationary state sets in for \( \tau <t<\tau _{L} \) .
For \( T>T_{c} \), this state is close to the equilibrium one. However, for
\( T<T_{c} \) the intense losses of the particles from the condensate due to
spin relaxation determine the appearance of the quasistationary state with the
compensating flux of particles into the region of low energies. As turns out,
the distribution function for the non-condensate fraction of the gas in such
state differs drastically from the equilibrium distribution function. An important
feature of this ``steady-flux'' distribution is a sharp increase of the number
of particles in the low-energy range of the spectrum. For the trap geometry,
this results in a strong increase of the density of normal fraction and therefore
causes the anomalous transition line shift for non-condensed atoms. Note that
the attempt to explain the anomalous increase of the non-condensate density
by introducing a rather artificial model of the formation of metastable dense
droplets has recently been undertaken in \cite{f}.

We confine ourselves to the time scale of \( \tau \ll t\ll \tau _{L} \) and
suppose that the quasistationary regime sets in. In order to simplify the analysis
we will consider a gas in a spherically symmetric harmonical potential with
frequency \( \omega  \). We suppose that the condensate is in the so-called
Thomas-Fermi regime at \( T<T_{c} \), that is, \( n_{c}(0)U_{0}\gg \hbar \omega  \)
where \( U_{0}=4\pi \hbar ^{2}a/m \) and \( m \) is the atom mass. This means
that the condensate density is determined by the expression
\begin{equation}
\label{TF}
n_{c}(r)=\frac{2\mu _{c}-m\omega ^{2}r^{2}}{2U_{0}}
\end{equation}
valid for \( r<r_{\mu } \) where \( r_{\mu }=(2\mu _{c}/m\omega ^{2})^{1/2} \)
and \( \mu _{c}=n_{c}(0)U_{0} \) is the chemical potential of the condensate. 

At \( T<T_{c} \) the loss of the particles is determined mostly by the condensate
\begin{equation}
\label{ncdot}
\dot{n}_{c}(r)=-\frac{1}{2}L^{(2)}n_{c}^{2}(r).
\end{equation}
Here \( L^{(2)} \) is the spin relaxation coefficient determined for a normal
gas and the factor of \( 1/2 \) appears from the two-particle density correlator
for a condensate. Note that, in spite of the spatially inhomogeneous loss of
Bose-condensed particles, the condensate retains Thomas-Fermi density profile.
The point is that the redistribution of particles in the condensate, induced
by the mean-field interaction, is much faster than their loss. Indeed, the characteristic
evolution time in the condensate is \( \tau =\hbar /n_{c}(r)U_{0} \), while
the characteristic time of the particle loss is \( \tau _{l}=1/L^{(2)}n_{c}(0) \).
In the atomic hydrogen \( \tau _{l}/\tau \sim 10^{5} \)\emph{.} 

With the condensate density distribution (\ref{TF}) we find for the total particle
loss rate
\begin{equation}
\label{Q}
Q=\frac{16\pi }{105}L^{(2)}n_{c}^{2}(0)r_{\mu }^{3}.
\end{equation}
Let us consider the kinetic equation for the normal fraction in a harmonic potential.
We are interested in particles with energies \( \epsilon >\mu \gg \hbar \omega  \).
In this case we can disregard the discrete structure of the trap energy levels.
Under the assumption of ergodicity the quantum Boltzmann equation acquires the
form 
\begin{equation}
\label{KE}
\dot{n}(\epsilon _{1})=I_{coll}(\epsilon _{1},[n]).
\end{equation}
where \( n(\epsilon ) \) is the statistical average of occupation numbers.
Within the energy interval where the occupation numbers are large

\begin{eqnarray}
I_{coll} & = & W\int _{0}^{\infty }d\epsilon _{2}d\epsilon _{3}d\epsilon _{4}\chi (\epsilon _{1},\epsilon _{2},\epsilon _{3},\epsilon _{4})\delta (\epsilon _{1}+\epsilon _{2}-\epsilon _{3}-\epsilon _{4})\nonumber \\
 &  & \{n(\epsilon _{1})n(\epsilon _{2})[n(\epsilon _{3})+n(\epsilon _{4})]\nonumber \\
 &  & -n(\epsilon _{3})n(\epsilon _{4})[n(\epsilon _{1})+n(\epsilon _{2})]\}\label{CI} 
\end{eqnarray}

where
\begin{equation}
\label{hi}
\chi (\epsilon _{1},\epsilon _{2},\epsilon _{3},\epsilon _{4})=\left[ \min \left( 1,\frac{\epsilon _{2}}{\epsilon _{1}},\frac{\epsilon _{3}}{\epsilon _{1}},\frac{\epsilon _{4}}{\epsilon _{1}}\right) \right] ^{2},
\end{equation}
\begin{equation}
\label{W}
W=\frac{4ma^{2}}{\pi \hbar ^{3}}.
\end{equation}
 This kinetic equation differs from that for a homogeneous case (see, e.g.,
\cite{10,b}) only by the exponent in expression (\ref{hi}) (see, e.g., \cite{c,d}),
which is characteristic for a harmonic potential. As we will see below, the
increase of spatial density, accompanying BEC, comes from the energy region
where \( n(\epsilon )\gg 1 \). We took the latter condition into account explicitly,
deriving Eq. (\ref{CI}). 

The density in the energy space can be written as \( \rho _{\epsilon }=g(\epsilon )n(\epsilon ) \)
where \( g(\epsilon )=\epsilon ^{2}/2(\hbar \omega )^{3} \) is the density
of states in the harmonic potential. We can rewrite Eq. (\ref{KE}) in the form
of the continuity equation for the energy space \emph{
\[
\frac{\partial \rho _{\epsilon }}{\partial t}=-\frac{\partial }{\partial \epsilon }P(\epsilon )\]
}with the flux \emph{}\( P(\epsilon ) \)
\begin{equation}
\label{P}
P(\epsilon )=\int ^{\epsilon }d\epsilon 'g(\epsilon ')I_{coll}(\epsilon ',[n]).
\end{equation}
In the quasistationary regime \( P(\epsilon )=Q=const \). This means that the
flux (\ref{P}) has the same value for an arbitrary \( \epsilon  \). Taking
into account Eq. (\ref{CI}), this requirement can be fulfilled if the distribution
function has the form 
\begin{equation}
\label{n_e}
n(\epsilon )=\frac{A}{\tilde{\epsilon }^{\gamma }}
\end{equation}
where \( \tilde{\epsilon }=\epsilon /\hbar \omega  \). Let us substitute this
distribution function into Eq. (\ref{CI}). Introducing dimensionless variables
\( \xi _{i}=\epsilon _{i}/\epsilon _{1} \), after simple transformations we
find\emph{
\[
I_{coll}=W(\hbar \omega )^{2}A^{3}I_{0}(\gamma )\tilde{\epsilon }_{1}^{2-3\gamma }\]
} where\emph{
\begin{eqnarray}
I_{0}(\gamma ) & = & \int _{0}^{\infty }d\xi _{2}d\xi _{3}d\xi _{4}\chi (\xi _{2},\xi _{3},\xi _{4})\delta (1+\xi _{2}-\xi _{3}-\xi _{4})\nonumber \\
 &  & (\xi _{4}^{\gamma }+\xi _{3}^{\gamma }-\xi _{2}^{\gamma }-1)(\xi _{2}\xi _{3}\xi _{4})^{-\gamma }\nonumber 
\end{eqnarray}
} Inserting these relations into Eq. (\ref{P}), we find for the flux 
\[
P(\epsilon )\propto \frac{I_{0}(\gamma )}{5-3\gamma }\epsilon ^{5-3\gamma }.\]
 The requirement of the independence of the flux on \( \epsilon  \) leads to
\begin{equation}
\label{gamma}
\gamma =\frac{5}{3}
\end{equation}
 and simultaneously to \( I_{0}(\gamma )=0 \). The last result is fairly evident:
\( I_{coll} \) is equal to zero in a stationary \emph{}case. At the same time
the derivative \( I_{0}'(\gamma ) \) is finite and provides the permanent flux
\( P=Q \) to the low energies region. \emph{}The obtained results are in correspondence
with the general analysis of Zakharov (see \cite{Zakh} and also \cite{Svist}),
showing that the equation \( I_{coll}=0 \) has a nontrivial solution relevant
to a steady-state particle flux in the energy space.

For the relation between \( A \) and \( Q \), we obtain
\begin{equation}
\label{A}
A=C\left( \frac{Q}{W(\hbar \omega )^{2}}\right) ^{1/3}.
\end{equation}
One can estimate the dimensionless numerical coefficient \( C \) using the
total number of non-condensate particles \( N_{nc} \) if one assumes approximately
that Eqs. (\ref{n_e}),(\ref{gamma}) remain valid up to the maximum energy. 

It is worth to note that the numerical calculations of the BEC kinetics demonstrate
the formation of the distribution of the form (\ref{n_e}) (with \( \gamma >1 \))
before the real growth of condensate starts (see, e. g., \cite{b,e}). 

Using \( n(\epsilon ) \) from Eq. (\ref{n_e}), one can find the spatial density
distribution \emph{\( n(r) \)} for the normal fraction. The direct numerical
calculation shows that the main contribution to \( n(r) \) comes from the energy
levels \( \mu \ll \epsilon <T \). These levels are weakly affected by the presence
of the condensate (the volume occupied by the wavefunctions of a particle is
much larger than the condensate volume) and by the interparticle interaction
in the thermal cloud. This means that within a reasonable approximation we can
neglect the renormalization of the levels and use oscillator vavefunctions \emph{\( \psi _{\mathbf{p}}(\mathbf{r}) \)}
where \( {\bf p}\equiv (n_{r},l,m) \). The energy region \( (\epsilon -\mu )\sim \mu  \),
where the modification of the level structure is noticeable, gives a negligible
contribution to the spacial density, and we ignore this modification. Under
the assumption of ergodicity
\begin{equation}
\label{n_expan}
n({\bf r})=\sum _{\mathbf{p}}|\psi _{\mathbf{p}}({\bf r})|^{2}n(\epsilon _{\mathbf{p}})
\end{equation}
Since the spacial density of the thermal cloud is weakly sensitive to the low
energies, \emph{}at the presence of the condensate we simply truncate \emph{}the
sum (\ref{n_expan}) by the condition \( \epsilon _{\mathbf{p}}\geq \mu _{c} \). 

Since the concentration of the condensate fraction is confined to several percents,
the main part of the particles at \( T<T_{c} \) is in the non-condensate fraction
and \( N_{nc}\approx N \). At \( T>T_{c} \) the distribution function for
low energies at the actual absence of the flux compensating losses has an equilibrium
form \( n(\epsilon )=T/(\epsilon +|\mu |) \). Comparing this distribution with
Eqs. (\ref{n_e},\ref{gamma}), one can see that the maximum density of the
non-condensate fraction, determined in the trap mostly by the low-energy region
in the presence of flux at \( T<T_{c} \), can exceed significantly the equilibrium
value of the density at \( T>T_{c} \). Note that in the presence of condensate
the thermal equilibrium is absent and therefore the temperature may be regarded
only as a characteristic energy scale for the system. 

For the Doppler-free two-photon excitation, the spectral distribution is determined
by the red shift of the absorption line. The shift is caused by the change of
the coupling (scattering length) of an excited atom compared with the atom in
the ground state. In the quasiclassical approximation the shift is proportional
to the local density of particles, coinciding with \( n(r) \) beyond the condensate
region. Thus, the appearance of a condensate, accompanied by a sharp growth
of \( n(r) \) for small \( r \), causes a significant increase of the shift
of the non-condensate spectral distribution. 

For an optically thin sample, in the local density approximation the Doppler-free
spectral distribution for the two-photon absorption is proportional to the density
distribution \( G(n) \). For the spherically symmetric configuration, this
distribution reads
\begin{equation}
\label{G}
G(n)=4\pi r^{2}(n)\left| \frac{dr(n)}{dn}\right| n
\end{equation}
where \( r(n) \) is determined from condition \( n(r)=n \). For the non-condensate
fraction, \( n(r) \) is determined by (\ref{n_expan}).

Below we present approximate quantitative estimations for the described picture
on the basis of the experimental data \cite{1,2}. Since in the experiments
the collision time \( \tau \sim 10^{-3} \)s is much less than the lifetime
of the system \( \tau _{L}\sim 5 \)s, the quasistationary regime sets in for
times \( \tau \ll t\ll \tau _{L} \) as explained above. The maximum condensate
density is found to be \( n_{c}(0)\approx 5\times 10^{15} \)cm\( ^{-3} \).
We will consider a spherically symmetric harmonic trap with \( \omega =(\omega _{\perp }^{2}\omega _{z})^{1/3}=3.4\times 10^{3} \)Hz.
Here \( \omega _{\perp } \) and \( \omega _{z} \) are the frequencies of the
cylindrical trap used in \cite{1,2}. For \( L^{(2)}=1.1\times 10^{15} \)cm\( ^{3} \)/s
and \( a=0.65{\rm \stackrel{\circ }{A}} \) \cite{2,12}, we find \( Q=1.9\times 10^{9} \)s\( ^{-1} \)
for the particle loss rate. In the Thomas-Fermi approximation, the number of
particles in the condensate for the given parameters is \( N_{c}\approx 1.2\times 10^{9} \).
The largest uncertainty in the experimental data \cite{1} is related to the
value of the relative condensate fraction \( \chi  \). Here we use some average
value of \( \chi \approx 0.07 \). Then the total number of particles is \( N\approx 1.7\times 10^{10} \).
The critical temperature of the BEC transition is \( T_{c}\approx 60\mu  \)K
in this case. 

Let us assume that non-condensate particles are concentrated within the energy
interval \( [\mu ,\epsilon ^{*}] \), having distribution (\ref{n_e}). Comparing
the energies of a system with the distribution (\ref{n_e}) and with the equilibrium
distribution at temperature \( T \), we find \( \epsilon ^{*}=4.7T_{c}(T/T_{c})^{4} \)
for the same number of particles in the both cases. The parameter \( \epsilon ^{*}\gg \mu  \).
The increase of the spatial density of non-condensed particles originates from
the occupation of the energy interval where \( \epsilon  \) is significantly
smaller than \( \epsilon ^{*} \). The results are only weakly sensitive to
the behavior of \( n(\epsilon ) \) near the upper bound of the energy spectrum.
Our approximate definition of \( \epsilon ^{*} \) is made with the aim to compare
the results below and above \( T_{c} \), assuming the same number of particles.
Taking into account the estimated values of \( N_{nc} \), \( Q \) and \( W \)
from Eq.(\ref{W}) we can determine the coefficients of \( A \) and \( C \)
in (\ref{gamma},\ref{A}). At \( T\approx 0.9T_{c} \) we obtain \( A\approx 2.9\times 10^{5} \)
and \( C\approx 2.5 \).

Using distribution function (\ref{n_e}) with the values found for \( A \)
and \( \epsilon ^{*} \), one can determine the spatial density distribution
(\ref{n_expan}) for the non-condensate fraction. 

\begin{figure}[tbp]

\epsfig{file=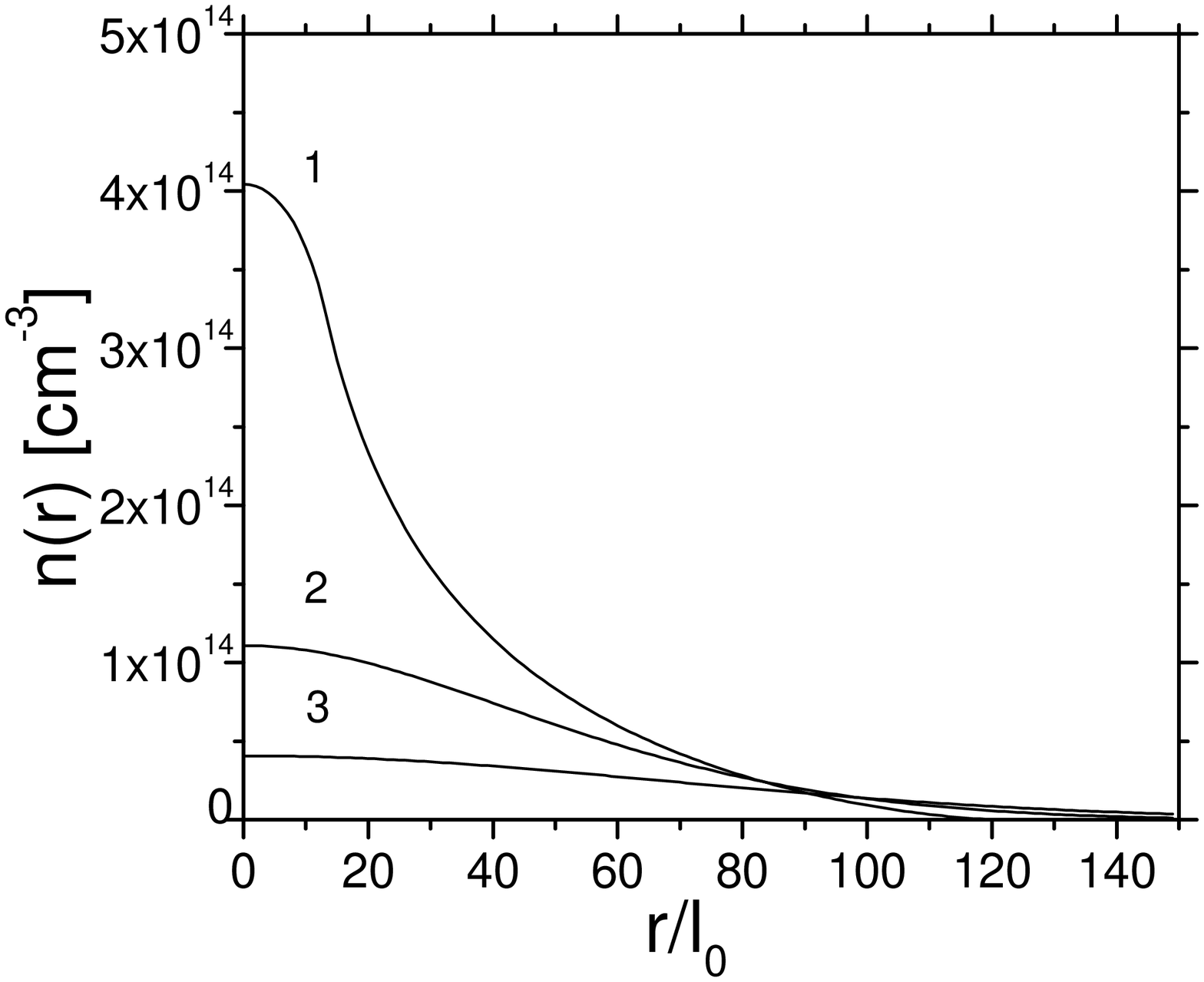, width=0.95\linewidth}

\caption{The spatial density distribution for the non-condensate fraction (\( N=1.7\times 10^{10} \) and \( T_{c}=60\mu  \)K): \( T_{1}=0.9T_{c} \) (curve 1), \( T_{2}=70\mu  \)K (curve 2) and \( T_{3}=120\mu  \)K (curve 3). }

\end{figure}

\begin{figure}[tbp]

\epsfig{file=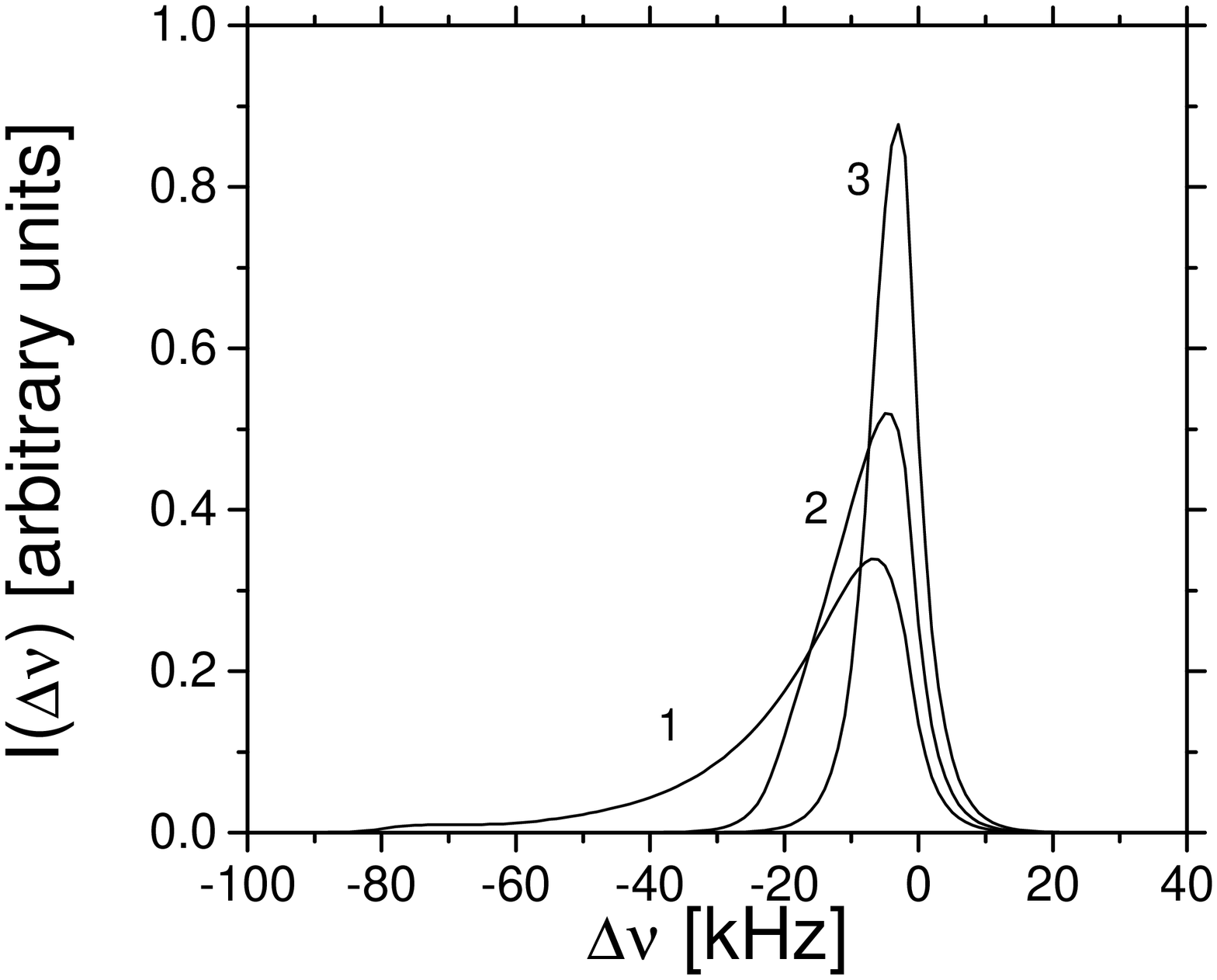, width=0.95\linewidth}

\caption{The spectral distribution of two-photon absorption for the non-condensate fraction. The parameters are the same as in FIG.1}

\end{figure}

The dependence \( n(r) \) is shown in FIG.1 (curve 1). For comparison, we present
the density profiles for two temperatures above \( T_{c} \): 70\( \mu  \)K
(curve 2) and 120\( \mu  \)K (curve 3). The length in the figure is given in
units of the characteristic trap size \( l_{0}=\sqrt{\hbar /m\omega } \). The
absence of condensate at \( T>T_{c} \) allows one to neglect the spin relaxation
losses and therefore to use the equilibrium distribution with the chemical potential
fixing the same number of particles.

Knowing \( n(r) \), we can find \( G(n) \) (\ref{G}) and therefore the spectral
distribution of two-photon absorption. To make a comparison with the experimental
results \cite{1}, we use the relation \( \Delta \nu (n)=\eta n \) for the
shift of absorption line and, according to \cite{1}, assume that the coefficient
\( \eta  \) has the same value both for the condensate and for the non-condensate
fraction. Regardless of a specific value of the scattering length for the interaction
between an excited atom and atoms in the ground state, one can use the relation
\( \Delta \nu (n)=\Delta \nu _{c}(n/n_{c}(0)) \). Here \( \Delta \nu _{c} \)
is the experimental value of the shift corresponding to the maximum condensate
density \( n_{c}(0) \). 

The dependence \( I(\Delta \nu ) \) for the density distributions presented
in FIG.1 is shown in FIG.2, \( \Delta \nu  \) being a shift of the laser frequency.
Calculating \( I(\Delta \nu ) \), we took into account the experimental linewidth
of \( \delta \nu \sim 3 \)kHz (see \cite{2}). 

The curves plotted in FIG.2 reproduce the characteristic behavior and scale
of the reconstruction observed in \cite{1} for the two-photon absorption spectrum
of the normal fraction with the appearance of a condensate. The nature of this
reconstruction lies in the formation of the special energy distribution of particles
due to the presence of the quasistationary flux of particles into the small
condensate region compensating the spin relaxation losses. 

Note in conclusion that all the presented calculations have an estimational
character due to adopted simplifications and uncertainty in a self-consistent
set of experimental data.

One of the authors (Yu.K.) acknowledges the fruitful discussions with T. J.
Greytak, D. Kleppner and L. Willmann. This work was supported by the Russian
Foundation for Basic Research (Grants 98-02-16262 and 99-02-18024) and by INTAS
(Grant INTAS-97-0972).

\end{document}